\pdfoutput=1
\documentclass[%
reprint,
showpacs,preprintnumbers,
 amsmath,amssymb,
 aps,
prb,
floatfix,
]{revtex4-1}

\usepackage{graphicx}
\usepackage{dcolumn}
\usepackage{bm}
\usepackage{hyperref}
\usepackage{multirow}

\RequirePackage{xspace}

\DeclareRobustCommand*{\etal}{et al.\xspace}

\newcommand{\SFO}%
{SrFe$_{12}$O$_{19}$}
\newcommand{\SFZSOx}%
{SrFe$_{12-x}$(Zn$_{0.5}$Sn$_{0.5}$)$_x$O$_{19}$}

\begin{document}

\title{Theory of magnetic enhancement in strontium hexaferrite through
  Zn-Sn pair substitution}

\author{Laalitha~S.~I.~Liyanage}
\affiliation{%
  Department of Physics and Astronomy, 
  Mississippi State University,
  Mississippi State, MS 39762, USA
}
\affiliation{%
  Center for Advanced Vehicular Systems, 
  Mississippi State University, 
  Mississippi State, MS 39762, USA
}

\author{Sungho~Kim}
\affiliation{%
  Center for Computational Sciences,
  Mississippi State University, 
  Mississippi State, MS 39762, USA
}

\author{Yang-Ki~Hong}
\affiliation{%
Department of Electrical and Computer Engineering and MINT Center, 
The University of Alabama, 
Tuscaloosa, AL 35487
}

\author{Ji-Hoon~Park}
\affiliation{%
Department of Electrical and Computer Engineering and MINT Center, 
The University of Alabama, 
Tuscaloosa, AL 35487
}

\author{Steven C. Erwin}
\affiliation{%
Center for Computational Materials Science, 
Naval Research Laboratory, 
Washington, D.C. 20375, USA
}

\author{Seong-Gon~Kim}\email{Author to whom correspondence should be addressed; kimsg@ccs.msstate.edu}
\affiliation{%
  Department of Physics and Astronomy, 
  Mississippi State University,
  Mississippi State, MS 39762, USA
}
\affiliation{%
  Center for Computational Sciences, 
  Mississippi State University, 
  Mississippi State, MS 39762, USA
}
\date{\today}

\begin{abstract}
  We study the site occupancy and magnetic properties of Zn-Sn
  substituted $M$-type Sr-hexaferrite
  SrFe$_{12-x}$(Zn$_{0.5}$Sn$_{0.5}$)$_x$O$_{19}$ with $x=1$ using
  first-principles total-energy calculations.  We find that in the
  lowest-energy configuration Zn$^{2+}$ and Sn$^{4+}$ ions
  preferentially occupy the $4f_1$ and $4f_2$ sites, respectively, in
  contrast to the model previously suggested by Ghasemi et al.
  [J. Appl. Phys, \textbf{107}, 09A734 (2010)], where Zn$^{2+}$ and
  Sn$^{4+}$ ions occupy the $2b$ and $4f_2$ sites.  Density-functional
  theory calculations show that our model has a lower total energy by
  more than 0.2~eV per unit cell compared to Ghasemi's model.  More
  importantly, the latter does not show an increase in saturation
  magnetization ($M_s$) compared to the pure $M$-type Sr-hexaferrite,
  in disagreement with the experiment.  On the other hand, our model
  correctly predicts a rapid increase in $M_s$ as well as a decrease
  in magnetic anisotropy compared to the pure $M$-type Sr-hexaferrite,
  consistent with experimental measurements.
\end{abstract}

\maketitle

Keywords: Site preference; Hexaferrite; Magnetic property

\section{Introduction}
\label{sec:intro} 

\begin{figure}[tpb]
  \centering
  \includegraphics[scale=0.5,keepaspectratio=true]{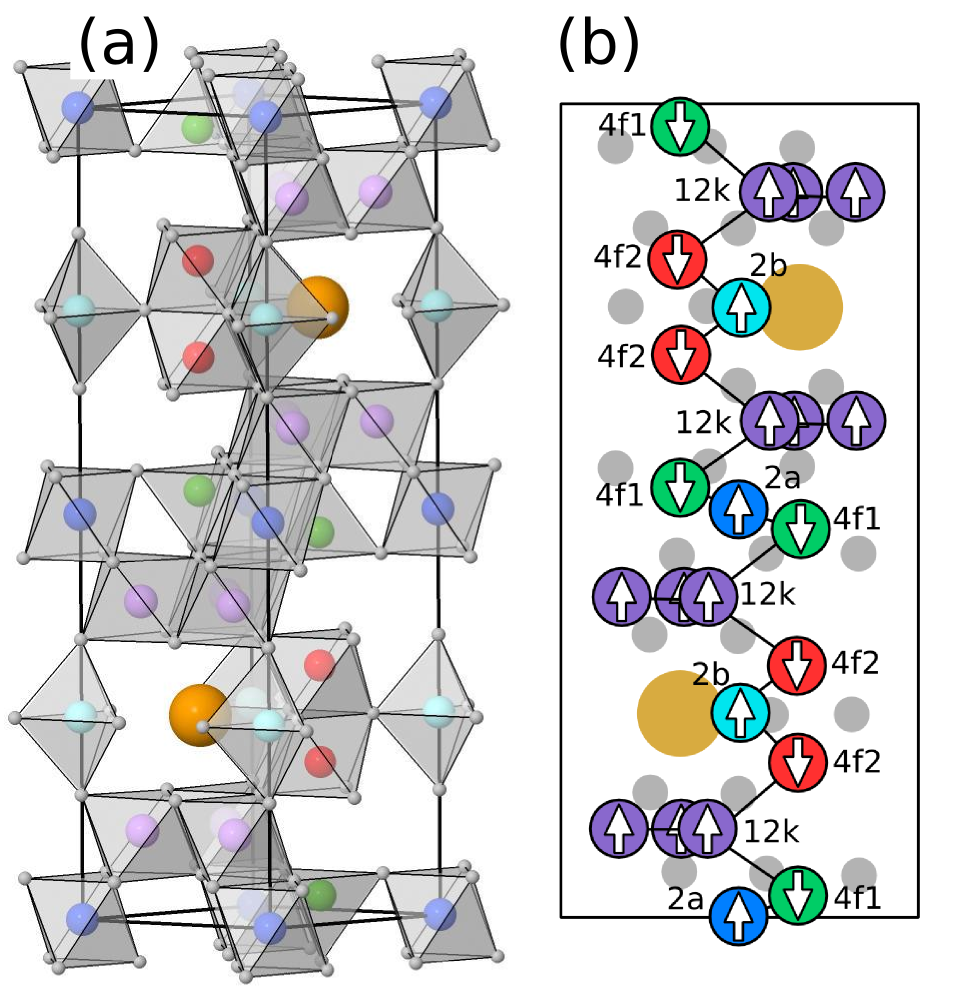}
  \caption{\label{fig:SFO}(Color online) (a) One double formula unit
    cell of \SFO. Two large gold spheres are Sr atoms and small gray
    spheres are O atoms.  Colored spheres enclosed by polyhedra formed
    by O atoms represent Fe$^{3+}$ ions in different inequivalent
    sites: $2a$ (blue), $2b$ (cyan), $12k$ (purple), $4f_1$ (green),
    and $4f_2$ (red). (b) A schematic diagram of the lowest-energy
    spin configuration of Fe$^{3+}$ ions of \SFO.  The arrows
    represent the local magnetic moment at each atomic site. }
\end{figure} 

Strontium hexaferrite \SFO\ (SFO) is widely used as a material for
permanent magnets, along with other $M$-type hexaferrites
$X$Fe$_{12}$O$_{19}$ ($X$ = Sr, Ba, Pb), due to their high Curie
temperatures, large saturation magnetization, high coercivity,
excellent chemical stability and low cost.\cite{Pang2010, Davoodi2011,
  Ashiq2012} As shown in Fig.~\ref{fig:SFO}, magnetism in SFO results
from Fe$^{3+}$ ions occupying five crystallographically inequivalent
sites in the unit cell: three octahedral sites ($2a$, $12k$, and
$4f_2$), one tetrahedral site ($4f_1$), and one trigonal bipyramid
site ($2b$).  SFO is a ferrimagnetic material that has 16 Fe$^{3+}$
ions with spins in the majority direction ($2a$, $2b$, and $12k$
sites) and 8 Fe$^{3+}$ ions with spins in the minority direction
($4f_1$ and $4f_2$ sites) as shown in
Fig.~\ref{fig:SFO}(c). Therefore, the substitution of nonmagnetic ions
into Fe sites with the minority spin direction has the potential to
increase the saturation magnetization of SFO by reducing the negative
contribution toward the saturation magnetization ($M_s$).
Consequently, investigations to improve the magnetic properties of
$M$-type hexaferrites have been made using various nonmagnetic
impurities.\cite{Asghar2012, Ashiq2012, Davoodi2011, Ashiq2011,
  Pang2010, Ghasemi2010, Iqbal2009, Bercoff2009, Nga2009, Ashiq2009,
  Iqbal2008, Rezlescu2008, Iqbal2007a, Iqbal2007, Qiao2007, Fang2005,
  Wang2004, Fang2004, Fang1999} Zr-Cd substituted SFO
(SrFe$_{12-x}$(Zr$_{0.5}$Cd$_{0.5}$)$_x$O$_{19}$) showed an increase
of $M_s$ up to $x=0.2$, while the coercivity decreased continuously
with increasing Zr-Cd concentration.\cite{Ashiq2009} For Er-Ni
substituted SFO, $M_s$ and coercivity steadily increased with Er-Ni
concentration.\cite{Ashiq2012} Substitution by Zn-Nb\cite{Fang2004},
Zn-Sn\cite{Ghasemi2010, Ghasemi2011} and Sn-Mg\cite{Davoodi2011}
increased $M_s$ and decreased coercivity.  These results call for a
systematic understanding, from first principles, of why certain
combinations of dopants lead to particular results. This theoretical
understanding will open the door to systematically optimizing the
properties of SFO.

There have been several previous first-principles investigations of
SFO.  Fang \etal\ investigated the electronic structure of SFO using
density-functional theory (DFT).\cite{Fang2003} Novak \etal\
calculated the electronic structure and exchange interactions of
barium hexaferrite using DFT.\cite{Novak2005} In spite of the
importance of substituted SFO, only a few theoretical investigations
have been done, and have focused on La
substitution.\cite{Kupferling2005, Novak2005a} To our knowledge, no
electronic structure calculation has been done on Zn-Sn-substituted
SFO.

In this work we use first-principles total-energy calculations to
study the site preference and magnetic properties of Zn-Sn substituted
$M$-type SFO \SFZSOx\ with $x=1$.  Based on total energy calculations,
we conclude that in the ground-state configuration Zn and Sn ions
preferentially occupy $4f_1$ and $4f_2$ sites, respectively.  This is
different from the model suggested by Ghasemi and
co-workers\cite{Ghasemi2010, Ghasemi2011}, where Zn and Sn ions occupy
$2b$ and $4f_2$ sites, respectively.  We further show that our model
predicts an increase of saturation magnetization as well as a
decrease in magnetic anisotropy energy (MAE) compared to the pure
$M$-type SFO ($x=0$) consistent with experimental observations.

\section{Methods}
\label{sec:method}

To determine the site preference of Zn and Sn atoms in Sr-hexaferrite,
we use first-principles total-energy calculations for configurations
of \SFZSOx\ at $x=0$ and $x=1$. We used a unit cell which contains two
formula units of SFO. The $x=1$ configuration was constructed by
substituting Zn and Sn, one atom each, into the Fe sublattices of
SFO. Total energies and forces were calculated using
density-functional theory (DFT) with projector augmented wave (PAW)
potentials as implemented in VASP.\cite{Kresse1996, Kresse1999} All
calculations were spin polarized according to the ferrimagnetic
ordering of Fe spins as shown in Fig.~\ref{fig:SFO}(b) as first
proposed by Gorter.\cite{Gorter1957, Fang2003} A plane-wave energy
cutoff of 520~eV was used for pure SFO and 400~eV for \SFZSOx\ with
$x=1$. Reciprocal space was sampled with a $7\times 7\times 1$
Monkhorst-Pack mesh\cite{Monkhorst1976} with a Fermi-level smearing of
0.2~eV applied through the Methfessel-Paxton
method\cite{Methfessel1989} for relaxations and the tetrahedron
method\cite{Bloch1994tetra} for static calculations. We performed
geometric optimizations to relax the positions of ions and cell shape
until the largest force component on any ion was less than
0.01~eV/\AA.  Electron exchange and correlation were treated with the
generalized gradient approximation (GGA) as parameterized by the
Perdew-Burke-Ernzerhof (PBE) scheme.\cite{Perdew1996}

To improve the description of localized Fe $3d$ electrons, we employed
the GGA+U method in the simplified rotationally invariant approach
described by Dudarev \etal.\cite{Dudarev1998} The method requires an
effective $U$ value ($U_{\text{eff}}$) equal to the difference between
the Hubbard parameter $U$ and the exchange parameter $J$.  The
effective $U$ (simply $U$ from now on) parameter can be determined in
a number of ways so as to reproduce various experimental or
theoretical results of particular interest. Here we choose the value
of $U$ that best reproduces the local magnetic moments of the Fe ions
obtained from a more rigorous calculation using the hybrid functional
of Heyd, Scuseria, and Ernzerhof (HSE).\cite{HSE2003, HSE2004,
  HSE2006} Thus our computational approach consisted of the following
sequence. (1) Use PBE to optimize the volume and internal coordinates
of pure SFO. (2) Use HSE to determine the individual Fe local moments
of pure SFO. (3) Within GGA+U, determine the value of $U$ that best
reproduces the HSE local moments. (4) Use GGA+U to investigate the
effects of varying the sites on which Zn and Sn substitute for Fe. 

\section{Results}
\label{sec:Results}

\subsection{Pure \SFO}

\begin{figure}[tpb]
  \centering
  \begin{tabular}{c}
    \includegraphics[scale=0.35,keepaspectratio=true]{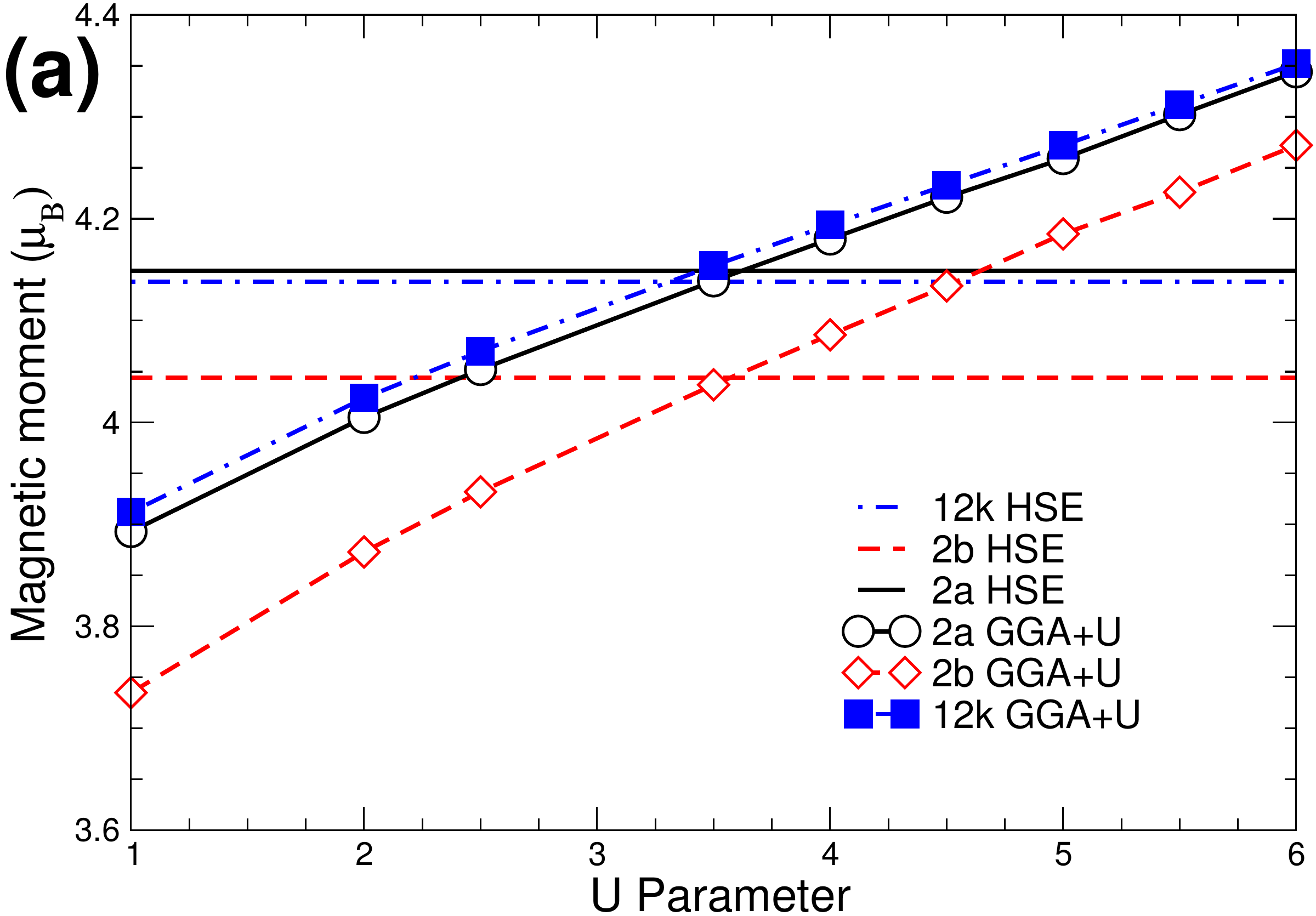} \\
    \includegraphics[scale=0.35,keepaspectratio=true]{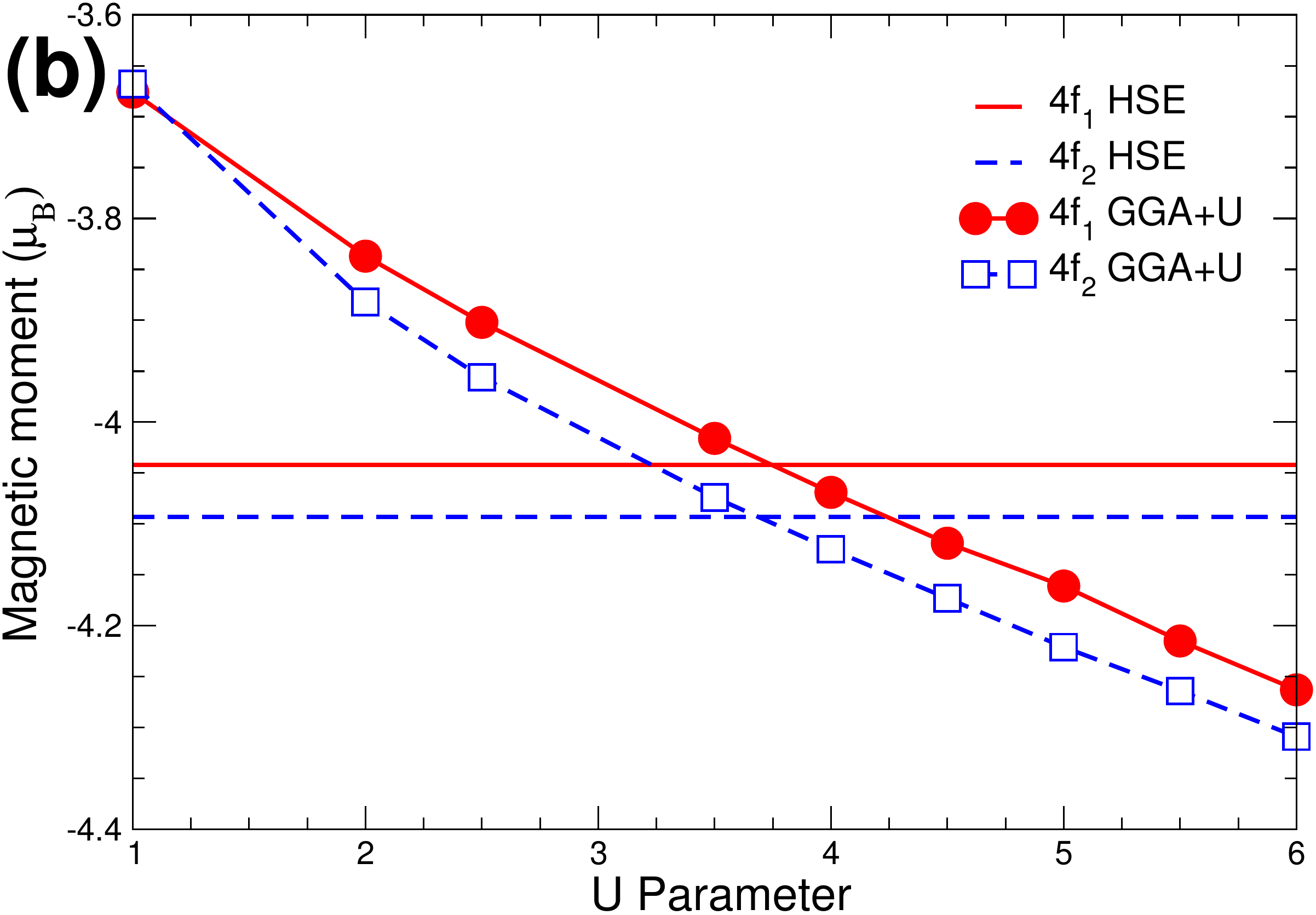} \\
  \end{tabular}
  \caption{\label{fig:magmom-Ueff} (Color online) Variation of local
    magnetic moments of the five inequivalent Fe sites of SFO with
    respect to the $U$ parameter; (a) positive magnetic moments for
    $2a$, $2b$, and $12k$ sites and (b) negative magnetic moments for
    $4f_1$ and $4f_2$ sites.  Values from HSE are shown as horizontal
    lines.}
\end{figure}

Pure SFO (\SFZSOx\ with $x=0$) has the hexagonal crystal structure and
the space group $P6_3/mmc$ as shown in Fig.~\ref{fig:SFO}. As shown in
Table~\ref{tab:SFO_GGAU}, the unit cell has 11 inequivalent sites: one
Sr site of multiplicity 2, five Fe sites of multiplicity 2, 2, 4, 4
and 12, and five oxygen sites of multiplicity 4, 4, 6, 12, and 12. In
total the double formula unit cell has 64 atomic sites. Our DFT
calculations showed that the spins of Fe$^{+3}$ ions on these sites
are arranged in an alternating fashion shown in Fig.~\ref{fig:SFO}(b)
as first proposed by Gorter.\cite{Gorter1957, Fang2003} This
antiferromagnetic arrangement of the spins of Fe$^{+3}$ ions in the
neighboring layers is consistent with the positive values of the
dominant intersublattice exchange integrals of
SrFe$_{12}$O$_{19}$\cite{Liyanage2014} and closely related
BaFe$_{12}$O$_{19}$\cite{Novak2005}.  DFT calculations showed that
deviation from this configuration always costs energy.\cite{Fang2003,
  Liyanage2014} Therefore, we set the initial spin direction of
Fe$^{+3}$ ions as Fig.~\ref{fig:SFO}(b) for all subsequent
calculations.

The dependence of the local magnetic moments of Fe$^{+3}$ ions on the
parameter $U$ is plotted in Fig.~\ref{fig:magmom-Ueff}.  The local
magnetic moments were computed using the PAW projector functions.
Values from the HSE calculations are also shown as horizontal lines
for comparison.  The local magnetic moments vary monotonically over
the range of $U$ values and, more importantly, they cross their
corresponding values of HSE calculation \textit{simultaneously} near
$U=3.7$~eV.  Therefore, we set $U$ to this value for all subsequent
calculations.  This value is between the two values, $U = 3.4$~eV and
$U = 6.9$~eV, chosen by Novak \etal\ for their GGA+U calculations of
Ba-hexaferrites with the full potential linearized augmented plane
wave (FPLAPW) method\cite{Novak2005}, and hence is quite reasonable.

\begin{table}[tpb]
  \caption{\label{tab:SFO_GGAU} Local magnetic moment of atoms in 
    different inequivalent sites of SFO calculated using GGA ($U = 0$), 
    GGA+U ($U=3.7$~eV), and HSE functionals. 
    Total magnetic moment ($m_{\text{tot}}$) of the unit cell is also given. 
    All moments are given in $\mu_B$.} 
  \begin{ruledtabular}
    \begin{tabular}{crrr}
      site & GGA & GGA+U & HSE \\
      \colrule  
      Sr ($2d$)    & -0.005 & -0.003 & -0.008 \\
      Fe ($2a$)    & 3.733  & 4.156  & 4.149  \\
      Fe ($2b$)    & 3.541  & 4.057  & 4.044  \\
      Fe ($4f_1$)  & -3.426 & -4.038 & -4.042 \\
      Fe ($4f_2$)  & -3.166 & -4.095 & -4.093 \\
      Fe ($12k$)   & 3.719 & 4.170 & 4.138  \\
      O ($4e$)     & 0.383 & 0.353 & 0.391  \\
      O ($4f$)     & 0.121 & 0.089 & 0.093  \\
      O ($6h$)     & 0.085 & 0.027 & 0.031  \\
      O ($12k$)    & 0.093 & 0.084 & 0.082  \\
      O ($12k$)    & 0.175 & 0.180 & 0.196  \\
      \hline
      $m_{\text{tot}}$ &  40  & 40 & 40 \\
    \end{tabular}
  \end{ruledtabular}
\end{table}

\begin{figure}[tpb]
  \centering
  \includegraphics[scale=0.35]{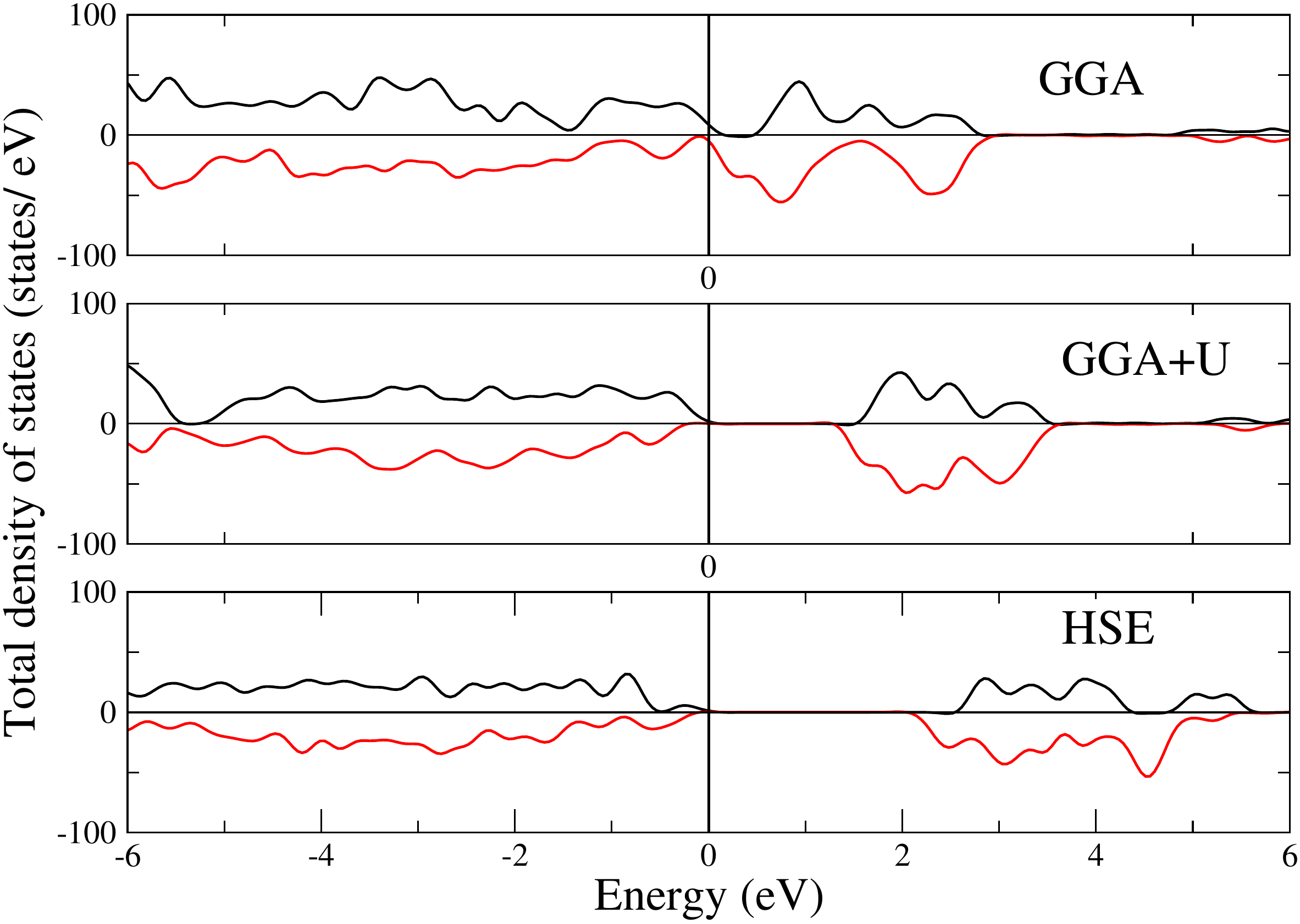} \\
  \caption{\label{fig:DOS-SFO} (Color online) Total densities of
    states (DOS) of \SFO\ calculated using GGA, GGA+U ($U=3.7$~eV),
    and HSE. Band gaps are 0.93 and 1.19~eV for GGA+U and HSE,
    respectively.  Positive (negative) values correspond to majority
    (minority) spin states.}
\end{figure}

For the optimized crystal structure we obtained the lattice constants
$a = 5.93$~\AA\ and $c = 23.20$~\AA\ in good agreement with the
experimental values of 5.88~\AA\ and 23.04~\AA,
respectively.\cite{Kimura1990} The total densities of states (DOS) for
SFO calculated with GGA ($U = 0$), GGA+U and HSE are shown in
Fig.~\ref{fig:DOS-SFO}.  HSE and GGA+U correctly predict an insulating
state while GGA predicts a metallic state. The local magnetic moments
of the 11 inequivalent Fe sites and the total magnetic moment are
given in Table~\ref{tab:SFO_GGAU}. Local magnetic moments of the Fe
sites predicted by GGA are much less than those by HSE. GGA+U's
prediction of the local magnetic moments is in good agreement with
HSE's results. The total magnetic moment values are 40~$\mu_B$ per
unit cell for all three methods.

\subsection{\SFZSOx\ with \ $x=1$}

\begin{figure}[tpb]
  \centering
  \includegraphics[scale=0.57,keepaspectratio=true]{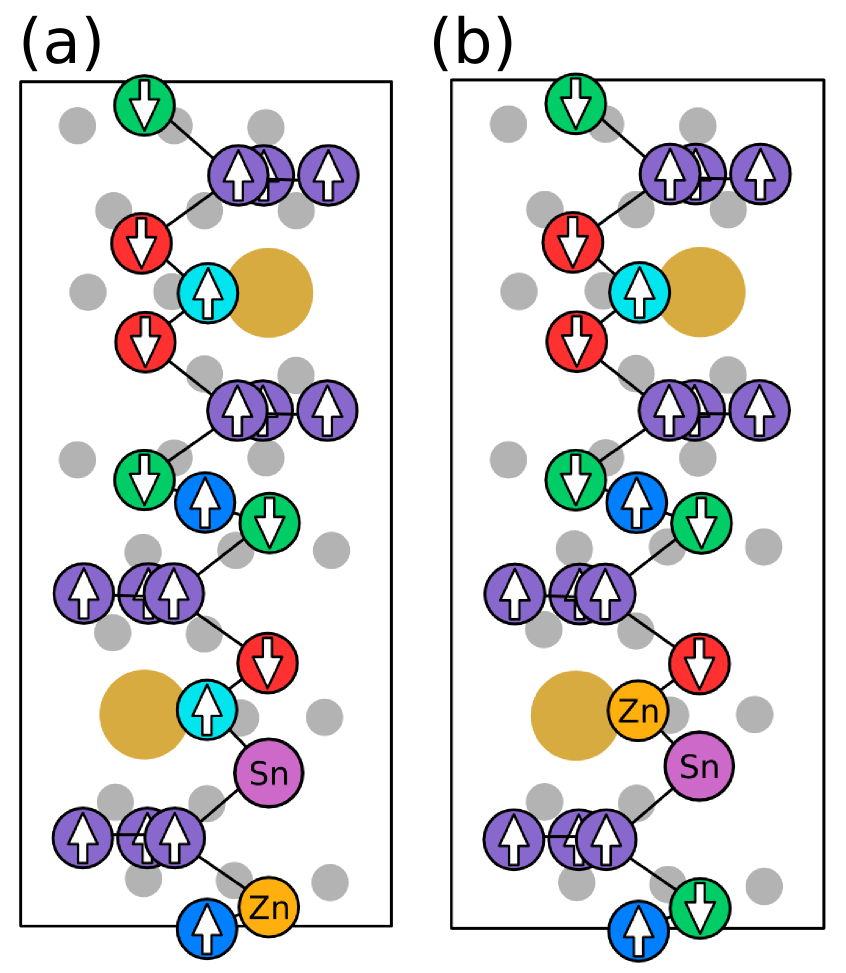}
  \caption{\label{fig:SFZSO}(Color online) The structures of \SFZSOx\
    with $x=1$ with spins oriented in easy axis (001): (a) Our
    proposed ground state configuration Zn($4f_1$)Sn($4f_2$).1 and (b)
    Ghasemi's proposed configuration Zn($2b$)Sn($4f_2$).1, which has
    higher total energy.  Zn and Sn atoms are labeled. Other atoms are
    colored as in Fig.~\ref{fig:SFO}.  }
\end{figure} 

For $x=1$, one Zn and one Sn atom are substituted at two of the 24 Fe
sites of unit cell as shown in Fig.~\ref{fig:SFZSO}. All Fe
sublattices listed in Table~\ref{tab:SFO_GGAU} have more than one
equivalent atomic site. Substitution of Zn-Sn atoms breaks the
symmetry of the equivalent sites of pure SFO.  Out of all 552
($=24\times 23$) possible structures, the 24 symmetry operations of
the space group $P6_3/mmc$ leave a total of 55 inequivalent
structures. We label these inequivalent configurations using the
convention of Zn(site for Zn)Sn(site for Sn).(unique index).  For
example, when Zn and Sn are substituted at the $2b$ and $12k$ sites,
respectively, there are 24
($=12\times 2$) possible structures but only two of them are
inequivalent. Therefore, they are labeled as Zn($2b$)Sn($12k$).1 and
Zn($2b$)Sn($12k$).2.  These structures were constructed by
substituting Zn and Sn atoms to the respective sites of a SFO unit
cell and optimizing all atomic positions using GGA+U.

\begin{table}[tpb]
  \caption{\label{tab:SFZSO} Nine lowest energy configurations of 
    \SFZSOx\ for $x=1$. Total energies 
    ($E_{\text{tot}}$) are relative to that of the ground state 
    configuration Zn($4f_1$)Sn($f_2$).1.  The total magnetic moment 
    ($m_{\text{tot}}$) and its change with respect to 
    pure SFO ($\Delta m_{\text{tot}}$) are also given.  
    All values are for a double formula unit cell 
    containing 64 atoms. 
  }
  \begin{ruledtabular}
    \begin{tabular}{ccccc}
      rank & configuration & $E_{\text{tot}}$ (eV) & $m_{\text{tot}}$ ($\mu_B$) 
      & $\Delta m_{\text{tot}}$ ($\mu_B$) \\
      \colrule
      1 & Zn($4f_1$)Sn($4f_2$).1 & 0.000 & 50 & 10 \\
      2 & Zn($4f_1$)Sn($4f_2$).2 & 0.133 & 50 & 10 \\
      3 & Zn($12k$)Sn($4f_2$).1 & 0.170 & 40 & 0 \\
      4 & Zn($4f_1$)Sn($12k$).1 & 0.188 & 40 & 0 \\
      5 & Zn($4f_1$)Sn($4f_2$).3 & 0.192 & 50 & 10 \\
      6 & Zn($2b$)Sn($4f_2$).1 & 0.211 & 40 & 0 \\
      7 & Zn($4f_1$)Sn($4f_2$).4 & 0.234 & 50 & 10 \\
      8 & Zn($4f_1$)Sn($2a$).1 & 0.262 & 40 & 0 \\
      9 & Zn($2a$)Sn($4f_2$).1 & 0.381 & 40 & 0 \\
    \end{tabular}
  \end{ruledtabular}
\end{table}
Table~\ref{tab:SFZSO} lists the nine lowest energy configurations. The
lowest-energy configuration Zn($4f_1$)Sn($4f_2$).1
(Fig.~\ref{fig:SFZSO}(a)) has Zn and Sn atoms substituted at the
$4f_1$ and $4f_2$ sites, respectively, and shows an increase of
10~$\mu_B$ per unit cell in $m_{\text{tot}}$ with respect to pure SFO
consistent with the experimental results of Ghasemi and
co-workers.\cite{Ghasemi2010, Ghasemi2011} On the other hand, the
structure suggested by Ghasemi with Zn$^{2+}$ in $2b$ site and
Sn$^{4+}$ in $4f_2$ site (configuration Zn($2b$)Sn($4f_2$).1 in
Table~\ref{tab:SFZSO} and shown in Fig.~\ref{fig:SFZSO}(b)) is
energetically less favorable by 0.211~eV.  Furthermore, this
configuration shows no increase in $m_{\text{tot}}$ contradicting the
experimental result that showed a rapid increase of saturation
magnetization with concentration $x$ ($\sim 7$\% increase at $x=1$).
We note that configurations with Zn and Sn substituted at the $4f_1$
and $4f_2$ sites show a considerable increase in $m_{\text{tot}}$
($=10$~$\mu_B$) while those at the $2a$, $2b$, and $12k$ sites show
zero increase.

In Table~\ref{tab:magmom-contrib} we compare the contribution of
different sublattices to the total magnetic moment in these two
proposed models for the structure of \SFZSOx\ with $x=1$. To see the
effect of Zn and Sn atoms in different sublattices, we split the
entries of sublattices containing these atoms ($4f_1$, $4f_2$, and
$2b$).  As expected, Zn$^{2+}$ and Sn$^{4+}$ ions carry negligible
magnetic moments regardless of their substitution sites.
Consequently, when they replace Fe$^{3+}$ ions in the minority spin
sites ($4f_1$ and $4f_2$), they eliminate a negative contribution and
hence increase the total magnetic moment per unit cell.  On the other
hand, when Zn$^{2+}$ and Sn$^{4+}$ ions replace Fe$^{3+}$ ions in the
majority spin site ($2b$), they eliminate a positive contribution and
hence reduce the total moment.  Therefore, configuration
Zn($4f_1$)Sn($4f_2$).1 results in an increase of 10~$\mu_B$ per unit
cell for the total magnetic moment $m_{\text{tot}}$, whereas
configuration Zn($2b$)Sn($4f_2$).1, where Zn and Sn are substituted to
the $2b$ and $4f_2$ sites, shows no increase in $m_{\text{tot}}$ as
the effect of Zn in the $2b$ site cancels out that of Sn in the $4f_2$
site.

\begin{table*}[tpb]
  \caption{\label{tab:magmom-contrib} Contribution of atoms in each 
    sublattice to the total magnetic moment of Zn-Sn-substituted 
    SFO structures 
    Zn($4f_1$)Sn($4f_2$).1 and Zn($2b$)Sn($4f_2$).1 compared with pure SFO.
    All moments are in $\mu_B$.  $\Delta m$ is measured relative to 
    the values for pure SFO. Note that the total magnetic moment of the 
    unit cell ($m_{\text{tot}}$) is slightly different than the sum of 
    local magnetic moments due to the contribution from the interstitial
    region.} 
  \begin{ruledtabular}
    \begin{tabular}{c|rr|rrr|rrr}
      \multicolumn{1}{c|}{\multirow{2}{*}{site}} & \multicolumn{2}{c|}{SFO} 
      & \multicolumn{3}{c|}{Zn($4f_1$)Sn($4f_2$).1} 
      & \multicolumn{3}{c}{Zn($2b$)Sn($4f_2$).1} \\
      \cline{2-9}
      & atoms & $m$ & atoms & $m$ & $\Delta m$ & atoms & $m$ & $\Delta m$ \\
      \colrule  
      $2d$ & 2 Sr & -0.01 & 2 Sr & 0.00 & 0.00 & 2 Sr & 0.00 & 0.00 \\
      $2a$ & 2 Fe &  8.31 & 2 Fe & 8.35 & 0.04 & 2 Fe & 8.30 & -0.01 \\
      \colrule
      \multirow{2}{*}{$2b$} 
      & 1 Fe & 4.06 & 1 Fe & 4.14 & 0.08 & 1 Fe & 4.05 & -0.01 \\
      & 1 Fe & 4.06 & 1 Fe & 4.05 & -0.01 & 1 \textbf{Zn} & -0.01 & -4.07 \\
      \colrule
      \multirow{2}{*}{$4f_1$} 
      & 3 Fe & -12.11 & 3 Fe & -12.06 & 0.03 & 3 Fe & -12.07 & 0.04 \\
      & 1 Fe & -4.04 & 1 \textbf{Zn} & 0.06 & 4.10 & 1 Fe & -4.03 & 0.01 \\
      \colrule
      \multirow{2}{*}{$4f_2$} 
      & 3 Fe & -12.29 & 3 Fe & -12.25 & 0.04 & 3 Fe & -12.30 & -0.02 \\
      & 1 Fe & -4.10 & 1 \textbf{Sn} & 0.06 & 4.16 & 1 \textbf{Sn} & 0.04 & 4.14 \\
      \colrule
      $12k$ & 12 Fe & 50.04 & 12 Fe & 50.30 & 0.26 & 12 Fe & 50.12 & 0.06 \\
      $4e$  & 4 O & 1.41 & 4 O & 1.43 & 0.02 & 4 O & 1.41 & 0.00 \\
      $4f$ & 4 O & 0.36 & 4 O & 0.57 & 0.21 & 4 O & 0.34 & -0.02 \\
      $6h$ & 6 O & 0.16 & 6 O & 0.41 & 0.25 & 6 O & -0.11 & -0.27 \\
      $12k$ & 12 O & 1.00 & 12 O & 1.62 & 0.62 & 12 O & 1.04 & 0.04 \\
      $12k$ & 12 O & 2.16 & 12 O & 2.15 & -0.01 & 12 O & 2.35 & 0.19 \\
      \colrule
      $\sum m$ & & 39.02 & & 48.81 & 9.79 & & 39.13 & 0.10 \\
      $m_{\text{tot}}$ & & 40 & & 50 & 10 & & 40 & 0 
    \end{tabular}
  \end{ruledtabular}
\end{table*}

\begin{table}[tpb]
  \caption{\label{tab:MAE} Magnetic anisotropy energy (MAE) and 
    magnetic anisotropy constant $K$ for pure SFO and two  
    configurations of \SFZSOx\ with $x=1$.}
  \begin{ruledtabular}
    \begin{tabular}{ccc}
      Configuration & MAE (meV) & $K$ (kJ$\cdot$m$^{-3}$) \\
      \colrule
      SFO & 0.84 & 190 \\
      Zn($4f_1$)Sn($4f_2$).1 & 0.45 & 100 \\
      Zn($2b$)Sn($4f_2$).1 & 0.53 & 120 
    \end{tabular}
  \end{ruledtabular}
\end{table}
In addition to an increase in the total magnetic moment, Ghasemi
\etal\ also reported a rapid decrease in anisotropy of SFO due to Zn
and Sn substitution.\cite{Ghasemi2010} This decrease was attributed to
the substitution of Zn$^{2+}$ ions on the $2b$ site. We calculated the
magnetic anisotropy energy (MAE) for the two models.  A
self-consistent calculation on the fully relaxed crystal structure was
performed. Then non-self-consistent calculations were performed with
the spin quantization axis oriented along the easy axis and the hard
plane.  MAE is the difference of total energies from these two
calculations:
\begin{equation*}
  E_{\text{MAE}} = E_{(100)} - E_{(001)}
\end{equation*} 
where $E_{(100)}$ is the total energy with spin quantization axis in
the hard direction and $E_{(001)}$ the total energy with
spin quantization axis in the easy direction.  Using the MAE,
the magnetic anisotropy constant $K$ can be computed:
\begin{equation*}
  K = \frac{E_{\text{MAE}}}{V\sin^2\theta}
\end{equation*} 
where $V$ is the equilibrium volume of the unit cell and $\theta$ is
the angle between the two spin quantization axis orientations
(90$^\circ$ in the present case). Results of these calculations are
presented in Table~\ref{tab:MAE}, which shows that both models show a
rapid decrease in anisotropy consistent with the experimental results.

\section{Discussion}

The present work reveals a new structural model for Zn-Sn-substituted
SFO, Zn($4f_1$)Sn($4f_2$).1, which is different from the previously
proposed model, Zn($2b$)Sn($4f_2$).1, by Ghasemi
\etal.\cite{Ghasemi2010, Ghasemi2011} Three independent results
support this new model:
\begin{enumerate}
\item Energy: The new model has a significantly lower total energy per
  Zn-Sn pair, by 0.211~eV, compared to Ghasemi's model.
\item Magnetization: The new model exhibit an increase of 10~$\mu_B$
  in total magnetic moment consistent with the experiment. In
  contrast, Ghasemi's model shows no increase in total magnetic
  moment as Zn atoms in the $2b$ site cancel the contribution of Sn
  atoms in the $4f_2$ site.
\item Anisotropy: Both models are compatible with the experimental
  observation of rapid decrease in magnetic anisotropy energy.
\end{enumerate}
Ghasemi's model and our model agree on the site for Sn atoms ($4f_2$),
but differ on the site for Zn atoms ($4f_1$ for the present model and
$2b$ for Ghasemi's).  To understand the source of this discrepancy, we
briefly review the reasoning behind Ghasemi's model. 

Ghasemi and co-workers proposed their model on the basis of
M\"{o}ssbauer spectroscopy and magnetic anisotropy data.  The measured
${}^{57}\text{Fe}$ M\"{o}ssbauer spectrum of \SFZSOx\ was fitted by a
superposition of five magnetically split components (sextets)
corresponding to the $2a$, $2b$, $4f_1$, $4f_2$, and $12k$ sites.  The
relative intensity ratio of the five sextets were adjusted to obtain a
satisfactory fit to the measured spectrum. The resultant ratios
correspond to the occupancy of Fe$^{+3}$ ions in each site. For pure
SFO, the intensity ratios corresponding to the $2a$, $2b$, $4f_1$,
$4f_2$, and $12k$ sites would have been 1:1:2:2:6.  For \SFZSOx\
Ghasemi \etal\ obtained fitted intensity ratios 1:0:2:1:6, meaning
that the Fe in the $2b$ and $4f_2$ sites were replaced by Zn and Sn.
The assignment of Zn to the $2b$ site was made by analogy to behavior
in the related systems BaFe$_{12-x}$(Zr$_{0.5}$Zn$_{0.5}$)$_x$O$_{19}$
and LaZnFe$_{11}$O$_{19}$.\cite{Li:2000, Obradors:1985}

One possible source of the discrepancy between two models is that the
interpretation of M\"{o}ssbauer spectra is not
straightforward.\cite{Ghasemi2011} Competing sets of intensity ratios
may produce fits of similar qualities.  It would be interesting to see
if a re-fitting based on our present model would reproduce the
M\"{o}ssbauer spectrum.

\begin{table}[tpb]
  \caption{\label{tab:SFO-single} Relative total energy and magnetic moment 
    of SFO when Zn or Sn atoms are substituted into Fe sites separately.
    Total energies ($E_{\text{tot}}$) are relative to the lowest energy 
    for each case.  The total magnetic moment 
    ($m_{\text{tot}}$) and its change with respect to the
    pure SFO ($\Delta m_{\text{tot}}$) are also given.  
    All values are for a double formula unit cell 
    containing 64 atoms. 
}
  \begin{ruledtabular}
    \begin{tabular}{ccccc}
      atom & site & $E_{\text{tot}}$ (eV) & $m_{\text{tot}}$ ($\mu_B$)
      & $\Delta m_{\text{tot}}$ ($\mu_B$) \\
      \colrule
      Zn & $4f_1$ & 0.000 & 45 & $+5$ \\
      Zn & $2a$ & 0.312 & 35 & $-5$ \\
      Zn & $12k$ & 0.367 & 35 & $-5$ \\
      Zn & $2b$ & 0.821 & 35 & $-5$ \\
      Zn & $4f_2$ & 1.073 & 45 & $+5$ \\
      \colrule
      Sn & $4f_2$ & 0.000 & 43 & $+3$ \\
      Sn & $2b$ & 0.055 & 33 & $-7$ \\
      Sn & $12k$ & 0.222 & 33 & $-7$ \\
      Sn & $2a$ & 0.462 & 33 & $-7$ \\
      Sn & $4f_1$ & 0.770 & 43 & $+3$ \\
    \end{tabular}
  \end{ruledtabular}
\end{table}
To further elucidate the stability of Zn($4f_1$)Sn($4f_2$).1 for
Zn-Sn-substituted SFO, we investigated the site preference of Zn and
Sn atoms separately.  Table~\ref{tab:SFO-single} shows the relative
total energies and magnetic moment of SFO when Zn or Sn atoms are
substituted into Fe sites separately. Our result shows that Zn prefers
$4f_1$ site while Sn prefers $4f_2$ site even when they are
substituted separately.  These preferences can be also understood in
terms of the oxidation states of all the ions.  The tetrahedral $4f_1$
site is enclosed by four O atoms (three $12k$-O and one $4f$-O). Each
of these O atoms is shared by four other Fe sites.  Since O ions have
a $-2$ oxidation state, we can assign $-2$ [$= 4(1/4)(-2)$] as the
oxidation state of the $4f_1$ site and conclude that the $4f_1$ site
will prefer to host ions with a $+2$ oxidation state such as
Zn$^{+2}$.  Similarly, the octahedral $4f_2$ site is enclosed by six O
atoms (three $12k$-O and three $6h$-O). Each of these O atoms is
shared by three other Fe sites.  Thus, we can assign $-4$ [$=
6(1/3)(-2)$] as the oxidation state of the $4f_2$ site and conclude
that the $4f_2$ site will prefer to host ions with a $+4$ oxidation
state such as Sn$^{+4}$.  On the other hand, the hexahedral (trigonal
bipyramidal) $2b$ site is enclosed by five O atoms: three $6h$-O atoms
are shared by three other Fe sites while two $4e$-O atoms by four
other Fe sites.  Thus, we can assign $-3$ [$= 3(1/3)(-2) +
2(1/4)(-2)$] as the oxidation state of the $2b$ site and conclude that
the $2b$ site will prefer to host ions with a $+3$ oxidation state such
as Fe$^{+3}$.  This analysis supports the site preference of the
present model where Zn$^{+2}$ and Sn$^{+4}$ ions occupy $4f_1$ and
$4f_2$ sites, respectively.

\section{Conclusion}

Using first-principles total energy calculations based on density
functional theory, we obtained the ground state structure for
Zn-Sn-substituted SFO, \SFZSOx\ with $x=1$. We showed that Zn$^{2+}$
and Sn$^{4+}$ ions preferentially occupy $4f_1$ and $4f_2$ sites. The
stable structure derived from our calculations shows a rapid increase
in saturation magnetization and a significant decrease in
magnetocrystalline anisotropy with respect to pure SFO that is
consistent with experimental observations.  We showed that the
previously proposed model with Zn atom in the $2b$ is a higher-energy
configuration.  We also showed that the lowest-energy configuration
determined from the present work shows a rapid decrease in anisotropy
consistent with the experimental results.

\section{Acknowledgment}

This work was supported in part by the U.S. Department of Energy
ARPA-E REACT program under Award Number DE-AR0000189, the Center for
Advanced Vehicular Systems (CAVS), and the Center for Computational
Science (CCS) at Mississippi State University.  Funding for this
project was also provided by the Office of Naval Research (ONR)
through the Naval Research Laboratory's Basic Research Program.
Computer time allocation has been provided by the High Performance
Computing Collaboratory (HPC$^2$) at Mississippi State University.

\bibliographystyle{apsrev4-1}
\bibliography{SFO}
\end{document}